\documentclass{aa}
\usepackage{graphicx}
\usepackage{natbib,times}

 \def\wm2{\hbox{\,Wm$^{-2}$}}

 \def\far{\ifmmode{\phi_{\rm act}}\else $\phi_{\rm act}$\fi}

\begin{document}
%
   \title{Solar total irradiance in cycle 23}

   \author{N.\,A.~Krivova\inst{1}
           \and S.\,K.~Solanki\inst{1,2}
           \and W.~Schmutz\inst{3}
          }

   \offprints{N.\,A. Krivova}

   \institute{Max-Planck-Institut f\"ur Sonnensystemforschung, 
              Max-Planck-Str.~2, 37191 Katlenburg-Lindau, Germany\\
              \email{natalie@mps.mpg.de}
              \and
              School of Space Research, Kyung Hee University,
              Yongin, Gyeonggi 446-701, Korea\\
              \and
              Physikalisch-Meteorologisches Observatorium Davos,
              World Radiation Center, Switzerland
             }

   \date{Received  2010; accepted 2011}

   \abstract
   {The most recent minimum of solar activity was deeper and longer than the
previous two minima as evidenced by different proxies of solar activity.
This is also true for the total solar irradiance (TSI) according to the PMOD
composite.}
   {The apparently unusual behaviour of the TSI has been interpreted as
evidence against solar surface magnetism as the main driver of 
the secular change in the TSI.
We test claims that the evolution of the solar surface magnetic field does
not reproduce the observed TSI in cycle 23.}
   {We use sensitive, 60-minute averaged MDI magnetograms and
quasi-simultaneous continuum images as an input to our SATIRE-S model and
calculate the TSI variation over cycle 23, sampled roughly
twice-monthly.
The computed TSI is then compared to the PMOD composite of TSI measurements
and to the data from two individual instruments, SORCE/TIM and
UARS/ACRIM~II, that monitored the TSI during the declining phase of cycle~23
and over the previous minimum in 1996, respectively.}
   {Excellent agreement is found between the trends shown by the
model and almost all sets of measurements.
The only exception is the early, i.e. 1996 to 1998, PMOD data.
Whereas the agreement between the model and the PMOD composite over
the period 1999--2009 is almost perfect, the modelled TSI shows a steeper
increase between 1996 and 1999 than implied by the PMOD composite.
On the other hand, the steeper trend in the model agrees remarkably well
with the ACRIM~II data.
A closer look at the VIRGO data, that make the basis of the PMOD composite
after 1996, reveals that only one of the two VIRGO instruments, the PMO6V,
shows the shallower trend present in the composite, whereas the DIARAD
measurements indicate a steeper trend.}
   {Based on these results, we conclude that
(1) the sensitivity changes of the PMO6V radiometers within VIRGO during the
first two years have very likely not been correctly evaluated, and that
(2) the TSI variations over cycle~23 and the change in the TSI levels
between the minima in 1996 and 2008
are consistent with
the solar surface magnetism mechanism.}


\titlerunning{Solar total irradiance in cycle 23}
\authorrunning{Krivova et al.}

\maketitle

%

\section{Introduction}

Solar total (i.e. integrated over all wavelengths) irradiance is the main
external source of energy input into the Earth's atmosphere and is thus one of
the key parameters of climate models.
Total solar irradiance (TSI) has been measured by a series of partly
overlapping space-borne instruments since 1978 and was found to vary on
different time scales from minutes to decades
\citep{willson-hudson-88,willson-hudson-91,froehlich-2008b}.
These variations have generally been attributed to the evolution of
solar surface magnetic features, such as sunspots, faculae and the network
\citep{foukal-lean-88,froehlich-lean-97,fligge-et-al-2000a,%
preminger-et-al-2002,%
krivova-et-al-2003a,wenzler-et-al-2005a,wenzler-et-al-2006a}.
In particular, models of the TSI assuming that all variations are entirely
due to the evolution of the solar photospheric magnetic flux explain up to
90\% of all observed variations up to the middle of cycle~23
\citep{wenzler-et-al-2006a}.

The minimum in 2008 was more extended and deeper than the two previous
minima (in 1976 and 1986) as indicated by many different indices
\citep[e.g.,][]{didkovsky-et-al-2009,heber-et-al-2009,heelis-et-al-2009,%
wang-et-al-2009,solomon-et-al-2010},
including the TSI \citep[see, e.g.,][]{froehlich-2009}.
\citet{froehlich-2009} found a decrease in TSI of more than 0.2~Wm$^{-2}$
compared to the minimum in 1996, which is more than shown by other indices
relative to the corresponding cycle amplitudes (an exception is the open
magnetic field).
He interpreted this decrease as evidence against the magnetic origin of the
secular change in the TSI.

This engendered the analysis by \citet{steinhilber-2010}, who used MDI
magnetogram and intensity synoptic charts and a model sculpturing the
SATIRE-S in order to reconstruct the TSI between 1996 and 2010.
He found a fair agreement between the calculated TSI and the PMOD data
between 1996 and 2004 but failed to reproduce the measurements afterwards
and thus arrived at the conclusion that `the TSI observation cannot be
described by the evolving manifestation of solar surface magnetism as
obtained from MDI CR synoptic magnetograms and photograms'.
Among possible explanations he lists the uncertainty in the irradiance
observations, a change in the global photospheric temperature or the fact
that the evolution of the weak magnetic flux might have been
underrepresented in his model.
The latter is, indeed, possible since the weak changes in the
irradiance levels at minima are believed to be
driven by the changes in the weak `background' magnetic flux
\citep[e.g.,][]{solanki-et-al-2002,krivova-et-al-2007a}.
By setting the cut-off at 50~G, which almost eliminates this background
flux, and by leaving out the latitudes above $\pm 65^\circ$,
\citet{steinhilber-2010} basically models the evolution of the TSI due to
active regions only.

This is implicitly confirmed by the recent study by \citet{ball-et-al-2010}.
They ran the SATIRE-S model based on 5-min averaged MDI magnetograms and
compared the outcome of the model with the SORCE total and spectral
irradiance measurements as well as with the PMOD TSI composite for the
period 2003--2009.
A very good agreement between the model and both TIM/SORCE and PMOD data was
found, with the linear correlation coefficient lying
above 0.98 in both cases, but
the modelled variability on the rotational time scale was weaker
compared to observations during the minimum period in 2008.
Note that \citet{ball-et-al-2010} employed a $3-\sigma$ noise cut-off of
roughly 40\,G and the limb threshold of $\mu =0.1$ (i.e. leaving out only
heliocentric angles above approximately $84^\circ$).

Here, we use the SATIRE-S model and the MDI magnetograms and continuum
images in order to re-calculate solar irradiance variations between 1996 and
2009 and to check whether calculated changes agree with observations.
In order to make sure that we include as weak magnetic features as possible
with MDI data, we use 60-min averaged magnetograms with a mean noise level
of about 3.9~G.
We also extend the analysis of \citet{steinhilber-2010} by comparing the
model's output with different TSI time series and not just the PMOD
composite.

We describe our model and all the data we use in this work in
Sect.~\ref{data}, present the results in Sect.~\ref{results} and
sum up our conclusions in Sect.~\ref{concl}.


\section{Data and Model}
\label{data}

\subsection{SATIRE and MDI data}
\label{satire}

In order to calculate TSI variations over the period 1996--2009, we 
employ the SATIRE-S model \citep{krivova-et-al-2003a,krivova-et-al-2010a},
which assumes that all irradiance changes on the considered time scales
(i.e. days to the solar cycle) are entirely due to the evolution of the solar
surface magnetic field.
SoHO MDI magnetograms and continuum images \citep{scherrer-et-al-95}
are employed in order to retrieve
the information on the surface distribution of different magnetic structures
at a given time.
The current version of the model distinguishes four types of surface features:
the quiet Sun (surface free of magnetic field above the noise level),
sunspot umbrae, sunspot penumbrae and faculae/network.

In order to include the weakest possible magnetic features in the analysis,
we use MDI magnetograms averaged over 60 minutes.
Averaging over longer intervals than this
is problematic in practice, since sufficiently
long sequences of single magnetograms become incrementally rarer, and a
still longer integration is not necessarily beneficial, since intrinsic
evolution and peculiar motion of magnetic features leads to a smearing of
the signal \citep[cf.][]{krivova-solanki-2004a}.
The noise level in the final magnetograms is roughly 3.9\,G on average, but
depends somewhat on the position on the disc as found by
\citet{ortiz-et-al-2002} and \citet{ball-et-al-2010}.
Note that due to the new calibration of the MDI magnetograms this is about
50\% higher than found by \citet{krivova-solanki-2004a}.
We set the noise threshold at the 3-$\sigma$ level, which removes around
99\% of the noise.
We also leave out the limb pixels having $\mu<0.1$
\citep[cf.][]{ball-et-al-2010}.
The line of sight magnetogram signal is corrected for foreshortening by
assuming the magnetic field in faculae and the network to be vertical.

MDI continuum images suffer progressively from a flat field distortion,
which becomes increasingly serious after around 2003--2004.
Therefore, we correct the images, dividing them by the median filters
produced once or twice per carrington rotation and kindly provided by the
MDI team (J. Sommers, 2007 and T. Hoeksema, 2008, priv. comm.).
These are the filters that are used by the MDI team to produce
level~2 continuum photograms.
The very limited number of the available level~2 images was the main
argument against employing these data directly instead of level~1.5.

The continuum images are rotated to co-align them in time with the
corresponding magnetograms, typically recorded within an hour of each other.
The continuum images are used to identify sunspot pixels, which are then
excluded from the corresponding magnetograms.
The remaining magnetic signal above the noise threshold is considered to be
faculae and the network.

We have originally selected images in such a way that they are sampled
roughly every two weeks, whenever possible.
Several shorter periods (e.g., the Halloween storm at the end of
October 2003 or end of December 1996) were covered more frequently.
A significant fraction of the final averaged magnetograms, however,
turned out to be corrupted for different reasons and had to be rejected.
The final set includes 262 pairs of 60-min averaged magnetograms and
corresponding continuum images covering the period November 1996 to April
2009.

The time-independent brightness spectra of all components are computed
\citep{unruh-et-al-99} from the corresponding model atmospheres using the
ATLAS9 code described by \citet{kurucz-93}.
The model has one free parameter, $B_{\rm sat}$, which takes into account
saturation of contrast with increasing concentration of magnetic elements
\citep[as suggested by the work
of][]{solanki-stenflo-85,ortiz-et-al-2002,voegler-2005}.
It is determined from a comparison of the model outcome with measurements.
The SATIRE model has been extensively described in a number of earlier
papers \citep[e.g.,][]{krivova-et-al-2003a,wenzler-et-al-2004a,%
wenzler-et-al-2005a,wenzler-et-al-2006a,solanki-et-al-2005a}.

\subsection{TSI observations}
\label{tsi_obs}

A number of space-borne instruments have being monitoring the TSI since 1978.
Their measurements mostly overlap in time, which should in principal allow
their cross-calibration.
A construction of a composite record is, nevertheless, quite challenging
because of a few but crucial exceptions.
Thus three composites have been produced, one each by
\citet{froehlich-2008b}, called PMOD composite, and by ACRIM
\citep{willson-mordvinov-2003} and IRMB \citep{dewitte-et-al-2004} teams.

Here, we use the PMOD composite since (1) SATIRE-S was previously found to
agree best with this composite
\citep{wenzler-et-al-2009a,krivova-et-al-2009b}, and (2)
Fr\"ohlich's (\citeyear{froehlich-2009}) analysis of the TSI trend between
the minima in 1996 and 2008 is based on this composite.
We use version 6.2 of 0904, i.e. exactly the same 
as used by \citet{froehlich-2009}.
In the period after 1996, the PMOD composite is based on SoHO/VIRGO
measurements \citep{froehlich-2009}.
VIRGO data are discussed in more detail in Section~\ref{results}.

In addition, we use data from two individual experiments.
For the period since 2003, i.e. the declining phase of cycle~23, we also
consider data from the SORCE/TIM instrument \citep{kopp-et-al-2005b}, which
show a slightly weaker decrease in TSI over the period 2003--2009
\citep{froehlich-2009}.
We also use UARS/ACRIM~II data \citep{willson-97} over the period covering
the activity minimum in 1996.
Besides ERBS/ERBE, this is the only instrument that monitored solar
irradiance uninterruptedly from the declining phase of cycle~22 to around
the maximum of cycle 23.
We also considered ERBS/ERBE data, but they are significantly noisier.
Although they can possibly be used to trace long-term changes, on
shorter time scales (at least, up to a year) they show a significantly lower
correlation with other data sets, such as EURECA/SOVA~1 and SOVA~2 or
Nimbus~7/ERB than, e.g, the ACRIM~II data \citep{mecherikunnel-96}.
Thus although we have also tried to compare the results to ERBS/ERBE
measurements, this turned to be ineffective and we do not discuss them
here.


\section{Results}
\label{results}

\begin{figure*}[]
\centering
   \includegraphics[width=16.0cm]{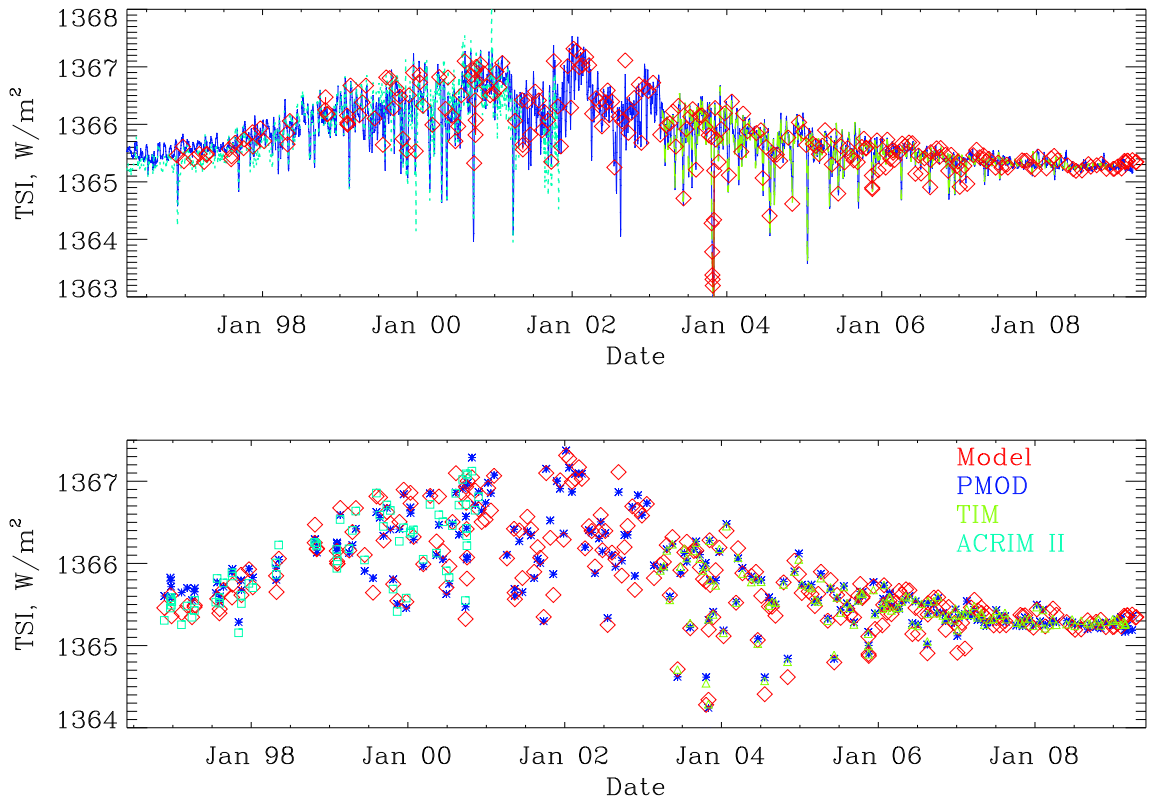}
\caption{Total solar irradiance over cycle 23: reconstructed using SATIRE-S
from 60-min averaged MDI magnetograms (red diamonds; consecutive days are
connected by red lines) and measured. Measurements by 2 instruments,
ACRIM~II (pine green) and TIM/SORCE (light green),
are shown as well as the PMOD composite of measurements (blue).
In the bottom panel, measurements are only shown on the days when the
SATIRE-S values based on 60-min MDI magnetograms are available.
}
\label{fig_cycle}
\end{figure*}

\begin{figure*}[]
\centering
   \includegraphics[width=16.0cm]{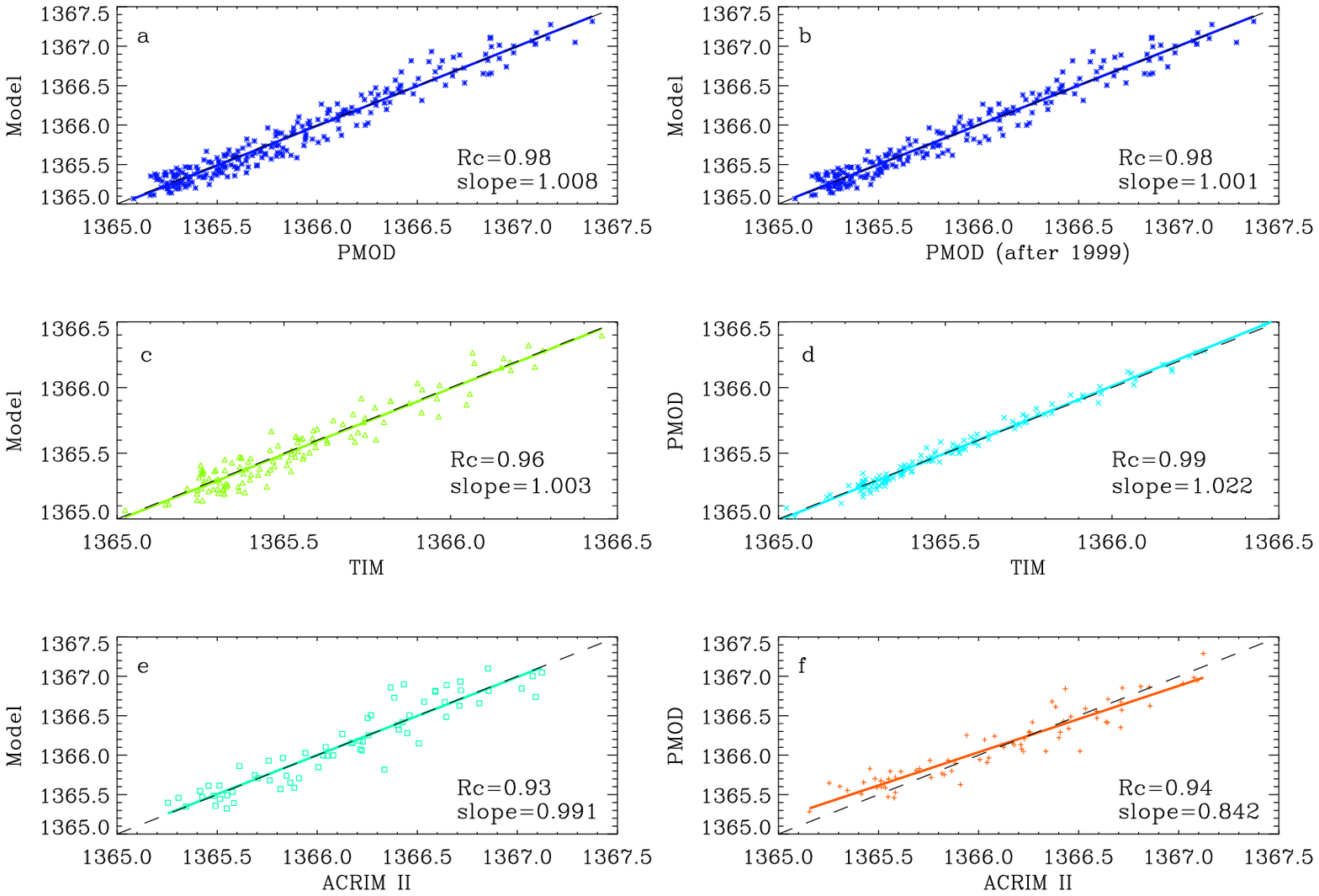}
\caption{
Comparison of the model with different data sets: {\bf a)} PMOD composite
(1996--2009);  {\bf b)} PMOD composite   
(1999--2009); {\bf c)} TIM;  {\bf e)} ACRIM~II, and of the PMOD composite
with the measurements by the 2 other individual instruments: {\bf d)} TIM
and {\bf f)} ACRIM~II.
The coloured straight lines show the slopes of the linear regressions
between the sets that are compared. The dashed black line shows the expected
ideal slope of 1.
In each panel, the corresponding correlation coefficient, $R_c$, and the
slope are indicated.
}
\label{fig_comp}
\end{figure*}

The reconstructed TSI is plotted in Fig.~\ref{fig_cycle} (red diamonds) and
is compared to the PMOD composite (blue), ACRIM~II (pine green) and TIM
(light green) measurements.
In the top panel all data are shown, whereas in the bottom panel only data
on the days when the model values are available, are kept, in order to
facilitate the comparison.
All data are shifted to fit the PMOD mean absolute level after 1999.
Note that the free parameter in the SATIRE-S model is fixed to achieve the
best agreement with the PMOD composite and is not varied further when
comparing SATIRE's output to other data sets.
The value of the free parameter is 452~G, which is very close to the value
of 450~G obtained by \citet{steinhilber-2010} using MDI synoptic
charts.
This is about 60\% higher than the value of 280~G found by
\citep{krivova-et-al-2003a} from a comparison with the level~2 VIRGO data
(version v5\_005\_0301), which is mainly due to the re-calibration of the
MDI magnetograms carried out by the MDI team \citep{tran-et-al-2005}.
It is easy to see that the model generally agrees quite well with all the
data sets, which is further confirmed by Fig.~\ref{fig_comp} and
Table~\ref{table}.

Figure~\ref{fig_comp} compares the model to each of the data sets directly:
PMOD (panels a and b), TIM (c) and ACRIM~II (e).
Also shown is the comparison of the PMOD data to those from the two
individual experiments, TIM (d) and ACRIM~II (f).
The coloured straight lines in Fig.~\ref{fig_comp} show the linear
regression fits between the sets compared in each panel, while the black
dashed lines (in most cases hidden behind the coloured lines) show the
`ideal' fits with the slope of 1.
Also listed in each panel are the corresponding correlation coefficients and
the slopes of the linear regression fits.
For this comparison, we have left out the data during big spot passages (i.e.
values below 1365~W/m$^2$), in order to avoid their strong effect on the
correlations and slopes since the main accent of this paper is the analysis
of the long-term changes in the TSI.

The correlation coefficients, their squares and the slopes of the linear
regression fits are summarised in Table~\ref{table}.
Note that the correlation coefficients emphasise the agreement on shorter time
scales (i.e. the scatter of the individual data points), whereas the slopes
of the linear regression better describe the long-term changes.
The best short-term agreement is reached between the TIM and PMOD data
($r_c=0.99$), although the correlation is only slightly lower for the
SATIRE-S vs. PMOD ($r_c=0.98$), and is still significantly high for
the SATIRE-S vs. TIM ($r_c=0.96$).
The correlation for both the SATIRE-S and PMOD vs. ACRIM~II is somewhat
weaker ($r_c=0.93$ and 0.94, respectively), implying that on shorter time
scales ACRIM~II is noisier than PMOD or TIM.

The slopes of the linear regressions between SATIRE-S and all data sets,
PMOD, TIM and ACRIM~II are all very close to 1 (the deviation is less than
1\% in all cases).
We emphasise that the free parameter of the model was fixed to provide the
best fit with the PMOD data and was not changed when comparing with the
other data sets, so that the slightly worse agreement (in the
slopes) with the other two sets is due to the model set up.

\begin{figure*}[]
\centering
   \includegraphics[width=14.0cm]{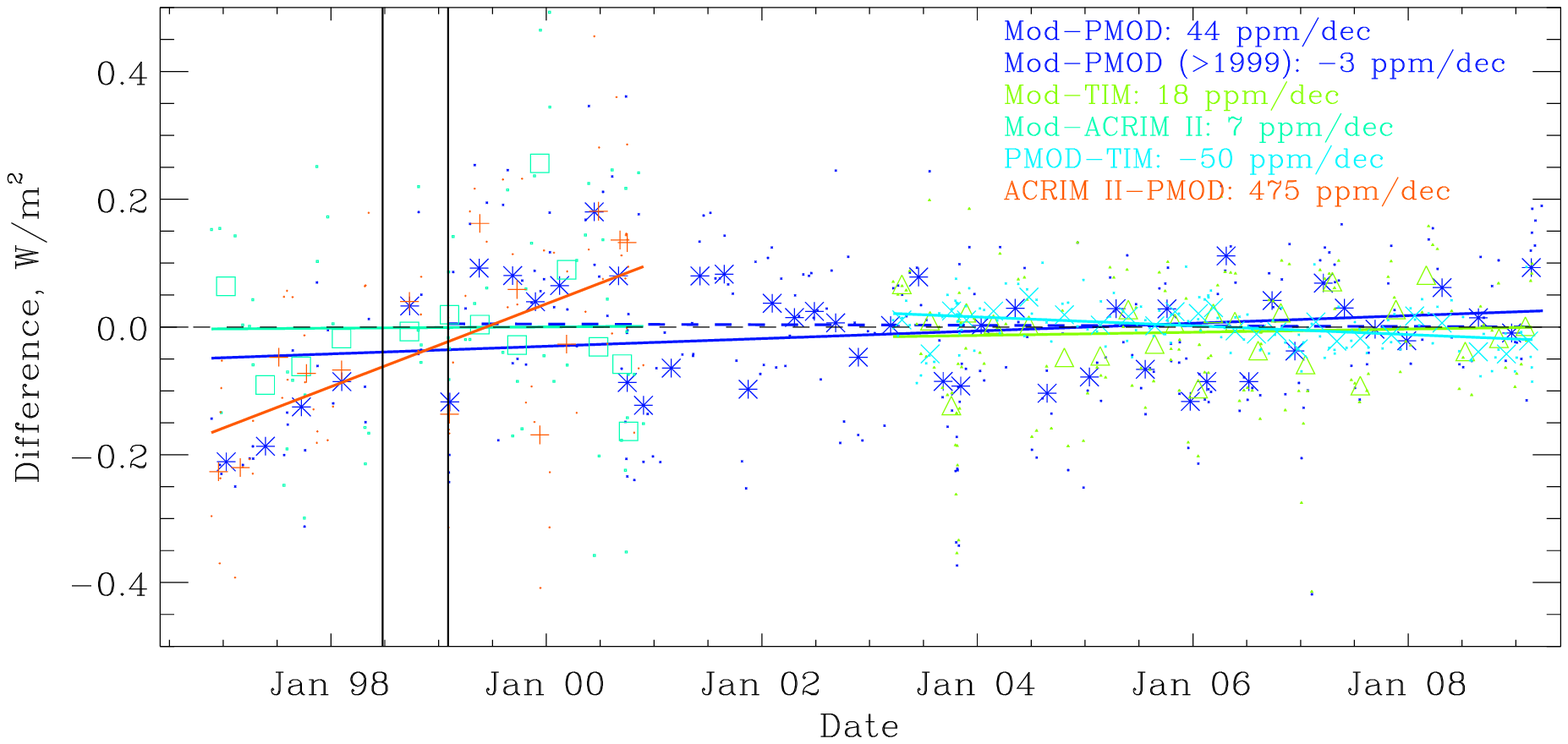}
\caption{
The difference, in W/m$^2$, between the model and the data: PMOD (blue),
ACRIM~II (pine green), TIM (light green),
and between the PMOD and the other two data sets: ACRIM~II (orange) and TIM
(cyan).
Dots show values for individual days, whereas bigger symbols represent bins
of 5 individual points.
The corresponding trends with time are indicated by the
straight lines and listed in the right top corner.
The vertical black lines show the beginning and the end dates of the
period without regular contact with SoHO.
}
\label{fig_trends}
\end{figure*}

A closer look at Fig.~\ref{fig_cycle} (see also Fig. ~\ref{fig_trends})
reveals, however, that the PMOD values lie slightly higher than
SATIRE-S and ACRIM~II values before approximately 1998/1999.
Note that ACRIM~II values were shifted to fit the PMOD level after the
recovery of SoHO in 1999.
Indeed, if PMOD values before 1999 are excluded, the slope of the linear
regression between SATIRE-S and PMOD becomes essentially 1 (see
Fig.~\ref{fig_comp} and Table~\ref{table}).
The remaining difference is smaller than the difference between PMOD and
TIM (slope 1.022).

Although the short-term agreement (i.e. $r_c$) between SATIRE-S and ACRIM~II
is worse than between SATIRE-S and PMOD or TIM, which is due to the higher
scatter in ACRIM~II data, the slope of the linear fit between SATIRE-S and
ACRIM~II is still very close to 1 (0.991; remember that the free
parameter, $B_{\rm sat}$, was fixed to get best agreement with PMOD so that
the obtained slope between SATIRE-S and ACRIM~II came out automatically
without further adjustments).
At the same time, the slope between PMOD and ACRIM~II is the only one that
strongly diverges from unity, being only 0.84 (see Fig.~\ref{fig_comp} and
Table~\ref{table}).

%
%

\begin{table*}
\caption[]{
Comparison between the model and different data sets and between the PMOD
composite and the two individual instruments.
}
\label{table}
\begin{tabular}{llccccr}
\hline
\hline
\noalign{\smallskip}
Series-1  & Series-2 & Period         & $r_c$ & $r^2_c$ & Slope\tablefootmark{a} 
                                                               & Trend\tablefootmark{a} \\
&&&&&& [ppm/dec] \\
\hline
\noalign{\smallskip}
Model     & PMOD     & 1996--2009     & 0.98  & 0.96    & 1.008$\pm 0.01$ & 44$\pm 17$  \\
Model     & PMOD-99  & 1999--2009     & 0.98  & 0.96    & 1.001$\pm 0.01$ & -3$\pm 19$   \\
Model     & TIM      & 2003--2009     & 0.96  & 0.92    & 1.003$\pm 0.02$ & 18$\pm 34$   \\
PMOD      & TIM      & 2003--2009     & 0.99  & 0.98    & 1.022$\pm 0.01$ & -50$\pm 14$  \\
Model     & ACRIM~II & 1996--2000\tablefootmark{b} 
                                      & 0.93  & 0.86    & 0.991$\pm 0.05$ &  7$\pm 135$   \\
PMOD      & ACRIM~II & 1996--2000\tablefootmark{b}\
                                      & 0.94  & 0.88    & 0.842$\pm 0.04$ & -475$\pm 101$ \\
\hline
\hline
\end{tabular}\\
\tablefoot{
Listed are: the data sets that
are compared; period over which they are compared;
the linear correlation coefficients and their squares, slopes
of the linear regression fits to the data sets, and the trends with time for
the difference between the two sets that are compared (Series-1~--
Series-2).}
\tablefoottext{a}{Estimates of the uncertainty are based on
the scatter of the measurements about the fitted lines plotted in
Figs.~\ref{fig_comp} and \ref{fig_trends}.}
\tablefoottext{b}{ACRIM~II data become very noisy after around 2000.}
\end{table*}

In Fig.~\ref{fig_trends}, we plot the difference
between different pairs of irradiance sets.
The dots signify differences on individual days, whereas the bigger symbols
show bins of 5 days on which 60-min magnetograms are available (i.e. 5
dots).
The regression lines highlight the trends.
The slopes of the regressions are listed in the top right corner of
Fig.~\ref{fig_trends} and in the last column of Table~\ref{table}.
The difference between SATIRE-S and PMOD after 1999 displays a
negligible slope of -3~ppm/decade (which is less than $1\sigma$).
It is more than an order of magnitude higher
($>2\sigma$) if PMOD over the whole period
1996--2009 is considered, but most of this is obviously due to the different
trends before 1999.
Similarly, the differences between SATIRE-S (optimised to fit the PMOD
composite) and TIM or ACRIM~II also exhibit weak slopes of 18~ppm/decade and
7~ppm/decade, respectively (both below $1\sigma$).
The difference PMOD~-- TIM shows a stronger trend of about -50~ppm/decade,
as also found by \citet{froehlich-2009}.
Striking, however, is the very strong trend in the difference
ACRIM~II~-- PMOD.
With 475~ppm/decade, this is an order of magnitude stronger than the trend
for the differences PMOD~-- TIM or SATIRE-S~-- PMOD (if taken over the whole
period) and roughly two orders of magnitude stronger than the differences
between SATIRE-S~-- ACRIM~II or SATIRE-S~-- PMOD (if taken only since 1999).
Even with the higher uncertainty introduced by the bigger scatter in
ACRIM~II, the trend differs from zero by $4.7\sigma$.

The vertical black lines in Fig.~\ref{fig_trends} bound the period without
regular contact with SoHO.
The analysis carried out here  shows that the systematic difference between
SATIRE-S and PMOD composite is restricted to the period prior to the loss of
contact with SoHO.
Moreover, a comparison with ACRIM~II data shows that whereas SATIRE-S is
consistent with this data set, the PMOD composite is not.
Furthermore, a comparison between the data from the two instruments
composing SoHO/VIRGO, namely PMO6V and DIARAD, reveals a similar difference
in their trends in this period as shown by ACRIM~II or SATIRE-S vs. PMOD
composite.

From 1996 onwards, the PMOD composite is based on SoHO/VIRGO measurements.
VIRGO started operation in 1996, just around the cycle minimum and the
evaluation of the early trends in these data is not entirely free of
problems, since the two types of VIRGO radiometers, PMO6V and DIARAD, showed
very different behaviour
\citep[e.g.,][]{froehlich-2000,froehlich-finsterle-2001,froehlich-2009}.
This is further complicated by the interruption in SoHO measurements in
1998/1999.

\begin{figure}[]
\centering
   \includegraphics[width=9.0cm]{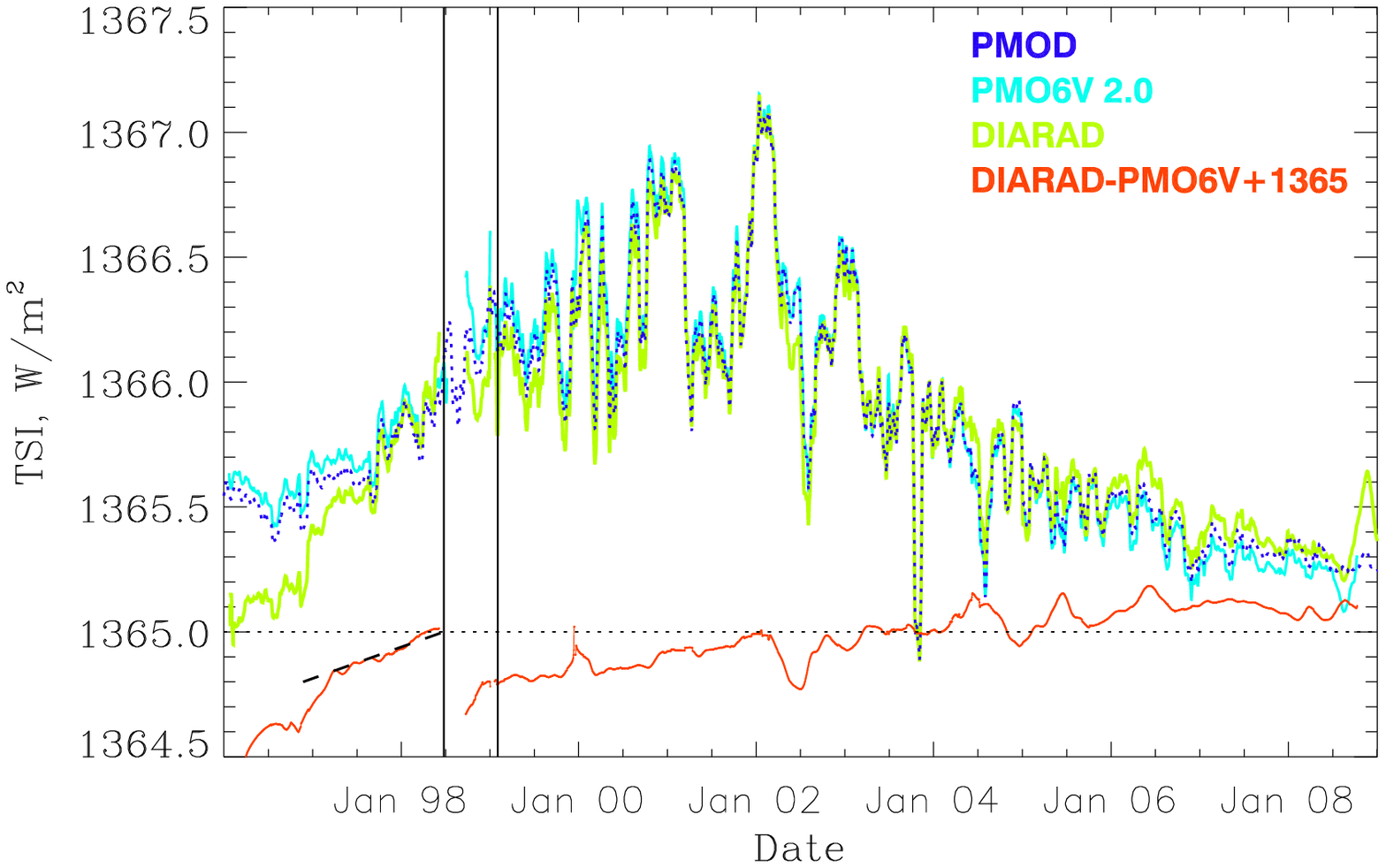}
\caption{ Monthly smoothed TSI measurements by PMO6V (light blue line),
DIARAD (light green line) and from the PMOD composite (dark blue dotted
line).
The PMO6V and DIARAD data were shifted to fit the mean PMOD level after
1999.
The red line shows the difference between DIARAD and PMO6V shifted by
1365~W/$m^2$.
The black dashed line marks the difference between SATIRE-S and PMOD in the
period before regular contact with SoHO was lost.
The vertical black lines mark the beginning and the end dates of the
period without regular contact with SoHO.
}
\label{fig_virgo}
\end{figure}

Figure~\ref{fig_virgo} shows the monthly smoothed data from
PMO6V\footnote{ftp://ftp.pmodwrc.ch/pub/data/irradiance/virgo/TSI/virgo\_tsi\_d\_v6\_002\_1009.dat}
(light blue line; level 2.0) and
DIARAD\footnote{http://remotesensing.oma.be/en/3361924-DiaradVIRGO.html}
\citep[light green line, level 2.0;][]{dewitte-et-al-2004a}
and their difference (red line) shifted by 1365~W/m$^2$.
Note that both data sets were shifted to fit the mean level of the PMOD
composite (dark blue dotted line) after 1999.
Obviously, in this earlier period the PMOD composite relies on the PMO6V
data, whereas the DIARAD data demonstrate a significantly steeper rise.
For the period around 1997--1998, this steeper rise is in good agreement
with ACRIM-II data and SATIRE-S results.
The dashed black line shows the mean difference between the model and the PMOD
data as determined by the points in Fig.~\ref{fig_trends} before the loss of
contact with SoHO.
We emphasise that since our set of 60-min MDI magnetograms starts only in
November 1996, we can argue neither pro nor contra the 
very steep increase in the DIARAD TSI in 1996 (which is not seen in ACRIM~II
data).

Cumulatively, these results suggest that the disagreement between the model
and the PMOD composite is more likely to lie with the early behaviour of
PMO6V data than with shortcomings of the SATIRE-S model.
Thus, the fact that the TSI level at the previous minimum is mismeasured by
roughly 0.2~W/m$^2$ in the PMOD composite (see Figs.~\ref{fig_trends} and
\ref{fig_virgo}), appears to be the main reason for the difference in the
behaviour of the TSI levels during the minima in 1996 and 2009 when compared
to other proxies as found by \citet{froehlich-2009}.

\section{Conclusions}
\label{concl}

We have reconstructed total solar irradiance using the SATIRE-S model based
on approximately twice-monthly 60-minute averaged MDI magnetograms and
continuum images.
We summarise our results here as follows:\\
1. We find
a good agreement between the model and the PMOD composite since
1999 and between the model and measurements of the two experiments:
SORCE/TIM and ACRIM~II, whose observations cover the declining
phase of cycle~23 and the activity minimum in 1996, respectively.
The slopes of the regression fits between the SATIRE-S and all sets of
observations all lie within 0.01 around 1.0, and the corresponding trends in
the differences SATIRE-S minus Observations are all below $\pm
50$~ppm/decade, i.e. less than the trend in the difference PMOD~-- TIM.
The agreement with PMOD increases significantly and becomes very
close to ideal within the error-bars (slope of the SATIRE-S vs.
PMOD fit is 1.001 and the trend for SATIRE-S~-- PMOD is -3~ppm/decade) if
PMOD data are only considered after the interruptions in SoHO operation in
1998/1999.\\
2. There is a strong trend (475~ppm/decade) in the difference between the
ACRIM~II and PMOD data in the period before the loss of regular contact with
SoHO, which roughly equals the trend in SATIRE-S--PMOD for this period.\\
3. A similar trend is also shown by the difference between the measurements
by DIARAD and PMO6V, the two VIRGO radiometers.
Whereas PMOD composite relies on PMO6V data, the TSI increase over 1997--1998
in our model is in good agreement with the trend shown by the DIARAD.

These results suggest that early degradation trends in PMO6V data
\citep{froehlich-2000,froehlich-finsterle-2001} might have not been fully
accounted for \citep[cf. also][]{froehlich-2009}.
For this reason,
the level of TSI during the activity minimum in 1996 seems to be overestimated in
the PMOD composite by roughly 0.2~W/m$^2$, which explains the apparently
different behaviour of the TSI over cycle~23 when compared to other proxies
as found by \citet{froehlich-2009}.
Therefore we conclude that the TSI variability in cycle~23 is fully
consistent with the solar surface magnetism mechanism
\citep{froehlich-lean-97,fligge-et-al-2000a,preminger-et-al-2002,%
krivova-et-al-2003a,wenzler-et-al-2006a}.

\begin{acknowledgements}
We are grateful to Jeneen Sommers, Todd Hoeksema and Phil Scherrer
for their advice and kind assistance in MDI data access and analysis.
We use data from VIRGO and MDI experiments on the cooperative ESA/NASA
mission SoHO.
We thank C.~Fr\"ohlich as well as SORCE/TIM, IRMB and ACRIM~II teams for
providing their data.
This work was supported by the
\emph{Deut\-sche For\-schungs\-ge\-mein\-schaft, DFG\/} project
number SO~711/1-3 and the WCU grant (No. R31-10016) funded by the Korean
Ministry of Education, Science and Technology.
\end{acknowledgements}

%
%

\begin{thebibliography}{40}
\expandafter\ifx\csname natexlab\endcsname\relax\def\natexlab#1{#1}\fi

\bibitem[{{Ball} {et~al.}(2011){Ball}, Unruh, Krivova, Solanki, \&
  Harder}]{ball-et-al-2010}
{Ball}, W.~T., Unruh, Y.~C., Krivova, N.~A., Solanki, S.~K., \& Harder, J.~W.
  2011, A\&A, submitted

\bibitem[{{Dewitte} {et~al.}(2004{\natexlab{a}}){Dewitte}, {Crommelynck}, \&
  {Joukoff}}]{dewitte-et-al-2004a}
{Dewitte}, S., {Crommelynck}, D., \& {Joukoff}, A. 2004{\natexlab{a}}, JGR
  (Space Physics), 109, 2102

\bibitem[{{Dewitte} {et~al.}(2004{\natexlab{b}}){Dewitte}, {Crommelynck},
  {Mekaoui}, \& {Joukoff}}]{dewitte-et-al-2004}
{Dewitte}, S., {Crommelynck}, D., {Mekaoui}, S., \& {Joukoff}, A.
  2004{\natexlab{b}}, Sol. Phys., 224, 209

\bibitem[{{Didkovsky} {et~al.}(2010){Didkovsky}, {Judge}, {Wieman}, \&
  {McMullin}}]{didkovsky-et-al-2009}
{Didkovsky}, L.~V., {Judge}, D.~L., {Wieman}, S.~R., \& {McMullin}, D. 2010, in
  ASP Conf. Ser., Vol. 428, SOHO-23: Understanding a Peculiar Solar Minimum,
  ed. {S.~R.~Cranmer, J.~T.~Hoeksema, \& J.~L.~Kohl}, 73--80

\bibitem[{{Fligge} {et~al.}(2000){Fligge}, {Solanki}, \&
  {Unruh}}]{fligge-et-al-2000a}
{Fligge}, M., {Solanki}, S.~K., \& {Unruh}, Y.~C. 2000, A\&A, 353, 380

\bibitem[{{Foukal} \& {Lean}(1988)}]{foukal-lean-88}
{Foukal}, P. \& {Lean}, J. 1988, ApJ, 328, 347

\bibitem[{{Fr{\" o}hlich} \& {Lean}(1997)}]{froehlich-lean-97}
{Fr{\" o}hlich}, C. \& {Lean}, J. 1997, ESA SP, 415, 227

\bibitem[{{Fr\"ohlich}(2000)}]{froehlich-2000}
{Fr\"ohlich}, C. 2000, Space Science Reviews, 94, 15

\bibitem[{{Fr{\"o}hlich}(2009)}]{froehlich-2009}
{Fr{\"o}hlich}, C. 2009, A\&A, 501, L27

\bibitem[{Fr{\"o}hlich(2009)}]{froehlich-2008b}
Fr{\"o}hlich, C. 2009, in {Climate and Weather of the Sun-Earth System
  (CAWSES): Selected Papers from the 2007 Kyoto Symposium}, ed. T.~T. et~al.
  (Setagaya-ku, Tokyo, Japan: Terra Publishing), 217--230

\bibitem[{{Fr\"ohlich} \& {Finsterle}(2001)}]{froehlich-finsterle-2001}
{Fr\"ohlich}, C. \& {Finsterle}, W. 2001, in Recent Insights Into the Physics
  of the Sun and Heliosphere~--- Highlights from SOHO and Other Space Missions,
  ed. P.~Brekke, B.~Fleck, \& J.~B. Gurman (ASP Conf. Ser., vol. 203), 105--110

\bibitem[{{Heber} {et~al.}(2009){Heber}, {Kopp}, {Gieseler},
  {M{\"u}ller-Mellin}, {Fichtner}, {Scherer}, {Potgieter}, \&
  {Ferreira}}]{heber-et-al-2009}
{Heber}, B., {Kopp}, A., {Gieseler}, J., {et~al.} 2009, ApJ, 699, 1956

\bibitem[{{Heelis} {et~al.}(2009){Heelis}, {Coley}, {Burrell}, {Hairston},
  {Earle}, {Perdue}, {Power}, {Harmon}, {Holt}, \&
  {Lippincott}}]{heelis-et-al-2009}
{Heelis}, R.~A., {Coley}, W.~R., {Burrell}, A.~G., {et~al.} 2009, GRL, 36,
doi:10.1029/2009GL038652

\bibitem[{{Kopp} {et~al.}(2005){Kopp}, {Lawrence}, \&
  {Rottman}}]{kopp-et-al-2005b}
{Kopp}, G., {Lawrence}, G., \& {Rottman}, G. 2005, Sol. Phys., 230, 129

\bibitem[{{Krivova} {et~al.}(2007){Krivova}, {Balmaceda}, \&
  {Solanki}}]{krivova-et-al-2007a}
{Krivova}, N.~A., {Balmaceda}, L., \& {Solanki}, S.~K. 2007, A\&A, 467, 335

\bibitem[{{Krivova} \& {Solanki}(2004)}]{krivova-solanki-2004a}
{Krivova}, N.~A. \& {Solanki}, S.~K. 2004, A\&A, 417, 1125

\bibitem[{{Krivova} {et~al.}(2003){Krivova}, Solanki, Fligge, \&
  Unruh}]{krivova-et-al-2003a}
{Krivova}, N.~A., Solanki, S.~K., Fligge, M., \& Unruh, Y.~C. 2003, A\&A, 399,
  L1

\bibitem[{{Krivova} {et~al.}(2011){Krivova}, {Solanki}, \&  
  Unruh}]{krivova-et-al-2010a}
{Krivova}, N.~A., {Solanki}, S.~K., \& Unruh, Y.~C. 2011, J. Atm. Sol.-Terr.
  Phys., 73, 223

\bibitem[{{Krivova} {et~al.}(2009){Krivova}, {Solanki}, \&
  Wenzler}]{krivova-et-al-2009b}
{Krivova}, N.~A., {Solanki}, S.~K., \& Wenzler, T. 2009, GRL, 36, L20101,
doi:10.1029/2009GL040707

\bibitem[{{Kurucz}(1993)}]{kurucz-93}
{Kurucz}, R. 1993, ATLAS9 Stellar Atmosphere Programs and 2 km/s grid.~Kurucz
  CD-ROM No.~13.~ Cambridge, Mass.: Smithsonian Astrophysical Observatory,
  1993., 13

\bibitem[{{Mecherikunnel}(1996)}]{mecherikunnel-96}
{Mecherikunnel}, A.~T. 1996, JGR, 101, 17073

\bibitem[{{Ortiz} {et~al.}(2002){Ortiz}, {Solanki}, Domingo, {Fligge}, \&
  Sanahuja}]{ortiz-et-al-2002}
{Ortiz}, A., {Solanki}, S.~K., Domingo, V., {Fligge}, M., \& Sanahuja, B. 2002,
  A\&A, 388, 1036

\bibitem[{{Preminger} {et~al.}(2002){Preminger}, {Walton}, \&
  {Chapman}}]{preminger-et-al-2002}
{Preminger}, D.~G., {Walton}, S.~R., \& {Chapman}, G.~A. 2002, JGR, 107 (A11),
  1354, DOI:10.1029/2001JA009169

\bibitem[{{Scherrer} {et~al.}(1995){Scherrer}, {Bogart}, {Bush}, {Hoeksema},
  {Kosovichev}, {Schou}, {Rosenberg}, {Springer}, {Tarbell}, {Title},
  {Wolfson}, {Zayer}, \& {MDI Engineering Team}}]{scherrer-et-al-95}
{Scherrer}, P.~H., {Bogart}, R.~S., {Bush}, R.~I., {et~al.} 1995, Sol. Phys., 162,
  129

\bibitem[{{Solanki} {et~al.}(2005){Solanki}, {Krivova}, \&
  Wenzler}]{solanki-et-al-2005a}
{Solanki}, S.~K., {Krivova}, N.~A., \& Wenzler, T. 2005, Adv. Space Res., 35,
  376

\bibitem[{{Solanki} {et~al.}(2002){Solanki}, {Sch{\"u}ssler}, \&
  {Fligge}}]{solanki-et-al-2002}
{Solanki}, S.~K., {Sch{\"u}ssler}, M., \& {Fligge}, M. 2002, A\&A, 383, 706

\bibitem[{{Solanki} \& {Stenflo}(1985)}]{solanki-stenflo-85}
{Solanki}, S.~K. \& {Stenflo}, J.~O. 1985, A\&A, 148, 123

\bibitem[{{Solomon} {et~al.}(2010){Solomon}, {Woods}, {Didkovsky}, {Emmert}, \&
  {Qian}}]{solomon-et-al-2010}
{Solomon}, S.~C., {Woods}, T.~N., {Didkovsky}, L.~V., {Emmert}, J.~T., \&
  {Qian}, L. 2010, GRL, 37, 16103, doi:10.1029/2010GL044468

\bibitem[{{Steinhilber}(2010)}]{steinhilber-2010}
{Steinhilber}, F. 2010, A\&A, 523, A39

\bibitem[{{Tran} {et~al.}(2005){Tran}, {Bertello}, {Ulrich}, \&
  {Evans}}]{tran-et-al-2005}
{Tran}, T., {Bertello}, L., {Ulrich}, R.~K., \& {Evans}, S. 2005, ApJ Suppl. Ser., 156,
  295

\bibitem[{{Unruh} {et~al.}(1999){Unruh}, {Solanki}, \&
  {Fligge}}]{unruh-et-al-99}
{Unruh}, Y.~C., {Solanki}, S.~K., \& {Fligge}, M. 1999, A\&A, 345, 635

\bibitem[{{V\"ogler}(2005)}]{voegler-2005}
{V\"ogler}, A. 2005, Mem. Soc. Astron. It., 76, 842

\bibitem[{{Wang} {et~al.}(2009){Wang}, {Robbrecht}, \&
  {Sheeley}}]{wang-et-al-2009}
{Wang}, Y., {Robbrecht}, E., \& {Sheeley}, N.~R. 2009, ApJ, 707, 1372

\bibitem[{{Wenzler} {et~al.}(2005){Wenzler}, {Solanki}, \&
  {Krivova}}]{wenzler-et-al-2005a}
{Wenzler}, T., {Solanki}, S.~K., \& {Krivova}, N.~A. 2005, A\&A, 432, 1057

\bibitem[{{Wenzler} {et~al.}(2009){Wenzler}, {Solanki}, \&
  {Krivova}}]{wenzler-et-al-2009a}
{Wenzler}, T., {Solanki}, S.~K., \& {Krivova}, N.~A. 2009, GRL, 36, L11102, 
doi:10.1029/2009GL037519

\bibitem[{{Wenzler} {et~al.}(2004){Wenzler}, {Solanki}, {Krivova}, \&
  {Fluri}}]{wenzler-et-al-2004a}
{Wenzler}, T., {Solanki}, S.~K., {Krivova}, N.~A., \& {Fluri}, D.~M. 2004,
  A\&A, 427, 1031

\bibitem[{{Wenzler} {et~al.}(2006){Wenzler}, {Solanki}, {Krivova}, \&
  {Fr\"ohlich}}]{wenzler-et-al-2006a}
{Wenzler}, T., {Solanki}, S.~K., {Krivova}, N.~A., \& {Fr\"ohlich}, C. 2006,
  A\&A, 460, 583

\bibitem[{Willson(1997)}]{willson-97}
Willson, R.~C. 1997, Science, 277, 1963

\bibitem[{{Willson} \& {Hudson}(1988)}]{willson-hudson-88}
{Willson}, R.~C. \& {Hudson}, H.~S. 1988, Nature, 332, 810

\bibitem[{{Willson} \& {Hudson}(1991)}]{willson-hudson-91}
{Willson}, R.~C. \& {Hudson}, H.~S. 1991, Nature, 351, 42

\bibitem[{{Willson} \& {Mordvinov}(2003)}]{willson-mordvinov-2003}
{Willson}, R.~C. \& {Mordvinov}, A.~V. 2003, GRL, 30, 1199, DOI
  10.1029/2002GL016038

\end{thebibliography}

\end{document}